%% file: deuteronKP.tex
\documentclass[ALICE,manyauthors,onecolumn]{cernphprep}
\usepackage[comma,square,numbers,sort&compress]{natbib}
\usepackage{hyperref}
\usepackage{footnote}
\makesavenoteenv{tabular}
\usepackage{lineno}
\usepackage{multirow}
\usepackage{amsmath}
\usepackage{siunitx}
\usepackage{xcolor}
\usepackage{wasysym}
\usepackage{placeins}
\usepackage{bm}
\usepackage{graphics,epsfig,graphicx}
\usepackage{amsbsy,gensymb}
\usepackage{amsfonts}
\usepackage{cancel}
\usepackage{braket}
\usepackage{isotope}
\usepackage{comment}
\usepackage[T1]{fontenc}
\usepackage{orcidlink}

\usepackage{stackengine}
\DeclareSIUnit\fm{\femto\meter}
\DeclareSIUnit\eVc{\eV\per\clight}
\DeclareSIUnit\keVc{\keV\per\clight}
\DeclareSIUnit\MeVc{\MeV\per\clight}
\DeclareSIUnit\MeVcc{\MeV\per\clight\squared}
\DeclareSIUnit\GeVc{\GeV\per\clight}
\DeclareSIUnit\GeVcc{\GeV\per\clight\squared}
\DeclareSIUnit\clight{\text{\ensuremath{c}}}
\DeclareSIUnit[number-unit-product = ]\percent{\char`\%}
\newcommand{\mycomment}[1]{}

\newcommand{\onethree}     {$\sqrt{s}~=~13$~Te\kern-.1emV\xspace}

\newcommand{\pt}{\ensuremath{p_{\rm T}}\,}
\newcommand{\mt}{\ensuremath{m_{\rm T}}\,}
\newcommand{\kt}{\ensuremath{k_{\rm T}}\,}

\newcommand{\ppNoSpace}{\ensuremath{\mathrm {p\kern-0.05em p}}}

\newcommand{\kd}{\ensuremath{\mbox{K$^{+}$--d}}~}
\newcommand{\pd}{\ensuremath{\mbox{p--d}}~}

\newcommand{\AkAd}{\ensuremath{\mbox{$\mbox{K}^{-}$--$\overline{\mbox{d}}$}}~}
\newcommand{\ApAd}{\ensuremath{\mbox{$\overline{\mbox{p}}$--$\overline{\mbox{d}}$}}~}
\newcommand{\kn}{\ensuremath{\mbox{K--N}}~}
\newcommand{\pp}{\ensuremath{\mbox{p--p}}~}
\newcommand{\pipi}{\ensuremath{\mbox{$\pi$--$\pi$}}~}

\newcommand{\kp}{\ensuremath{\mbox{K$^{+}$--p}}~}
\newcommand{\K}{$\mbox{K}^+$}
\newcommand{\Kbar}{$\mbox{K}^-$}
\newcommand{\pbar}{$\overline{\mbox{p}}$}
\newcommand{\dbar}{$\overline{\mbox{d}}$}
\newcommand{\plab}{\ensuremath{p}}

\def\bmx{{\bm x}}
\def\bmy{{\bm y}}

\newcommand{\pL}{\ensuremath{\mbox{p--$\Lambda$}}~}

\newcommand{\ks}           {\ensuremath{k^*}\xspace}
\newcommand{\rs}           {\ensuremath{r^*}\xspace}

\newcommand{\purityK}{\ensuremath{99.8}}
\newcommand{\purityAK}{\ensuremath{99.8}}
\newcommand{\purityp}{\ensuremath{98.2}}
\newcommand{\purityAp}{\ensuremath{97.9}}
\newcommand{\purityd}{\ensuremath{100}}
\newcommand{\purityAd}{\ensuremath{100}}

\newcommand{\MeVc}{\ensuremath{\mathrm{MeV}\kern-0.05em/\kern-0.02em \textit{c}}}
\newcommand{\GeVc}{\ensuremath{\mathrm{GeV}\kern-0.05em/\kern-0.02em \textit{c}}}
\newcommand{\GeVcSq}{\ensuremath{\mathrm{GeV}\kern-0.05em/\kern-0.02em \textit{c}^2}~}
\newcommand{\MeVcSq}{\ensuremath{\mathrm{MeV}\kern-0.05em/\kern-0.02em \textit{c}^2}~}

\newcommand{\LedLu}{Lednick\'y\xspace-Lyuboshitz\xspace}

\newcommand{\rKdcore}{\ensuremath{1.04\pm0.04}~}
\newcommand{\rpdcore}{\ensuremath{0.99\pm0.05}\,~}

\newcommand{\rNNcore}{\ensuremath{1.35\pm0.14}~}

\newcommand{\rKdeff}{\ensuremath{1.35^{+0.04}_{-0.05}}~}
\newcommand{\rKdeffoneg}{\ensuremath{1.22\pm0.04}\,~}
\newcommand{\rpdeff}{\ensuremath{1.08\pm0.06}\,~}
\newcommand{\rNNeff}{\ensuremath{1.43^{+0.16}_{-0.16}}\,~}

\newcommand{\events}{\ensuremath{1\times10^9}\,}

\newcolumntype{s}{>{\hsize=.5\hsize}X}

\begin{document}%

\begin{titlepage}
\PHyear{2023}
\PHnumber{181}      
\PHdate{24 August}  
\title{Exploring the strong interaction of three-body systems at the LHC}
\ShortTitle{}

\Collaboration{ALICE Collaboration\thanks{See Appendix~\ref{app:collab} for the list of Collaboration members}}
\begin{abstract}

Deuterons are atomic nuclei composed of a neutron and a proton held together by the strong interaction.
Unbound ensembles composed of a deuteron and a third nucleon have been investigated in the past using scattering experiments and they constitute a fundamental reference in nuclear physics to constrain nuclear interactions and the properties of nuclei.
In this work,
\kd and \pd femtoscopic correlations
measured  by the ALICE Collaboration in proton--proton (pp) collisions at $\sqrt{s}=13$~TeV at the Large Hadron Collider (LHC) are presented.
It is demonstrated that correlations in momentum space between deuterons and kaons or protons allow us to study three-hadron systems at distances comparable with the proton radius.
The analysis of the \kd correlation shows that the relative distances at which deuterons and protons or kaons are produced are around 2 fm.
The analysis of the \pd correlation shows that only a full three-body calculation that accounts for the internal structure of the deuteron can explain the data.
In particular, the sensitivity of the observable to the short-range part of the interaction is demonstrated.
These results indicate that correlations involving light nuclei in pp collisions at the LHC will also provide access to any three-body systems in the strange and charm sectors.

\end{abstract}
\end{titlepage}
\setcounter{page}{2}

\input{Content/1_Introduction.tex}
\input{Content/2_Led}

\input{Content/4_PisaProj}
\newenvironment{acknowledgement}{\relax}{\relax}
\begin{acknowledgement}
\section*{Acknowledgements}
The ALICE Collaboration is grateful to Professor Johann Haidenbauer and Professor Tetsuo Hyodo  for providing the derivation of the \kd scattering parameters, and to Stanis{\l}aw Mr\'owczy\'nski and Urs Wiedemann for fruitful discussions.
\input{fa_2023-07-31_Opt_C.tex}    
\end{acknowledgement}

\bibliographystyle{utphys}   
\bibliography{Content/Bibliography.bib}

\newpage
\appendix

\input{Content/7_Methods}

\newpage
\section{The ALICE Collaboration}
\label{app:collab}
\input{2023-07-31-Alice_Authorlist_2023-07-31_Opt_C.tex}  
\end{document}

%% file: Content/1_Introduction.tex

The study of multi-body systems is a key aspect of modern nuclear physics because of its relevance for the structure of nuclear bound states~\cite{Hammer:2012id,Hergert:2020bxy,Lynn2019} and for the equation of state of dense nuclear matter~\cite{Baldo:1996vg,Bombaci2}.
Effects that go beyond the simple addition of the strong interaction between pairs of nucleons emerge already in the description of the most basic properties of light nuclei.  
Realistic potentials describing nucleon--nucleon interactions~\cite{Wiringa:1994wb,Stoks:1994wp} have therefore been complemented with phenomenological models of three-nucleon forces~\cite{Pudliner:1997ck,Coon:2001pv}.
In calculations using potentials derived from chiral effective field theories (EFTs)~\cite{Epelbaum:2019kcf,Hergert:2020bxy}, the multi-body forces appear naturally as subleading terms in the chiral expansion. They find contributions of the order of 10\% to the ground-state energies of $A \leq 12$ nuclei from genuine three-body nucleon forces~\cite{LENPIC:2018ewt,Piarulli:2017dwd}, and deliver predictions for heavier and neutron-rich nuclear structures.

In this context, two-nucleon scattering data and properties of the $A=3$ systems are still the most important ingredients to constrain the parameters of nuclear interactions derived from EFT~\cite{Piarulli:2016vel,Reinert:2017usi,Saha:2022oep}.
In particular, differential scattering observables for the \pd system have allowed for the computation of a full-fledged three-body wave function that accounts for all the relevant two- and three-body interactions at work in the p--(pn) system for the short and the asymptotic ranges, providing an excellent description of the \pd scattering data~\cite{Kievsky:2001fq,PhysRevC.69.014002,Deltuva:2005xa}. 
In this work, it is demonstrated that such a standard candle of the three-body nuclear interaction can also be investigated by means of \pd correlations in momentum space measured at the LHC. 

Momentum correlations of the deuteron with other hadron species have already been considered as a tool to study both the deuteron production mechanism~\cite{Mrowczynski:2019yrr} and the final-state interactions for multi-body systems. Different experimental correlations such as \pd and d--d have been measured~\cite{Chitwood:1985zz,Pochodzalla:1986ova,Pochodzalla:1987zz,Wosinskapd} in O--Au reactions at $E/A$ $=$ 25, 35, and 60 MeV and in ${}^{40}$Ar--${}^{58}$Ni reactions at 77 MeV/u. The data showed a clear signature of the strong final-state interaction among light nuclei and nucleons, but the specific description of the process from a multi-body perspective, as well as the tools to precisely measure the relative distances between particles, were not available at the time of those analyses.

Deuteron--hadron momentum correlations can also be investigated at the LHC, since light (anti)nuclei can be abundantly produced and accurately measured in ultra-relativistic nucleus-nucleus collisions~\cite{NucleiCERNPS,ReviewAGS,RHIC1,RHIC2,RHIC6,nuclei_pp,nuclei_pp_PbPb,deuteron_pp_13TeV,deuteron_PbPb_276TeV}.
Recent femtoscopy analyses carried out by ALICE in pp, p--Pb, and Pb--Pb collisions have demonstrated that it is possible to study the strong interaction among several hadron pairs~\cite{ALICE:Run1,ALICE:pXi,ALICE:pK,ALICE:pOmega,ALICE:2021cpv,ALICE:2022enj,ALICE:2021szj} given the short distances at which hadrons can be produced in such colliding systems~\cite{Acharya:2020dfb}.

In this work, by means of a comprehensive study of the \kd and \pd correlation
functions measured in pp collisions at center-of-mass energy $\sqrt{s}=13$~TeV with the ALICE detector at the LHC, evidence is provided that deuterons are formed at average distances of the order of 2 fm from other hadrons.
The measurement of the \pd correlation at such short distances constitutes an innovative method to study three-body systems at the LHC, with the potential of extending such studies to the strangeness and charm sectors. 
Indeed, in the strangeness sector, a similar approach to the one adopted for standard nuclear physics is envisaged for the future by improving the database of hypernuclei~\cite{Feliciello:2015dua,Tolos:2020aln,Saito:2021gao}, scattering experiments~\cite{PhysRevLett.127.272303}, and femtoscopy measurements in hadron--hadron collisions~\cite{Fabbietti:2020bfg,alicereviewarxiv}. Direct measurements of the hyperon--deuteron systems at short distances would provide complementary information to these standard methods.

%% file: Content/2_Led.tex
\section*{Correlation function from scattering parameters}
Final-state interactions involving light nuclei such as deuterons have been studied in the past via scattering experiments~\cite{PhysRevLett.71.3762,PhysRev.72.662,BRUNE199813,HUTTEL1983435,CLEGG1995200}. In scattering theory, the nuclear interaction in the asymptotic regime can be investigated by associating a plane wave with the incoming particle and building the outgoing wave as a superposition of spherical waves with phase shifts $\delta_l$, with $l$ denoting the relative angular momentum between the projectile and the target. The interaction determines the values of $\delta_l$, and for $l=0$, referred to as $s$-wave scattering, a scattering length $a_0$ is commonly used to characterize the interaction at zero energy
and can be related to the differential cross section measured in scattering experiments.
These measurements also enable the determination of the effective range of the interaction $d_0$.

Scattering experiments have been already performed for the \kd and \pd systems allowing for the extraction of the corresponding scattering parameters, as reported in Table~\ref{tab:ScatteringLength_dp_kd}. In the case of \kd, such parameters are spin averaged, and they are calculated with two different methods: i) via an effective range fit (ER) to the cross section predictions at threshold anchored to the available scattering data~\cite{Takaki:2009ei}; ii) from the well-known \kn interactions~\cite{Aoki:2018wug} using the fixed-center approximation (FCA)~\cite{Kamalov:2000iy}.
The negative values of the \kd scattering length ($a_0$) refer to a repulsive strong interaction.
In the case of the p--d system, the parameters for the spin doublet ($S = 1/2$) and quartet ($S = 3/2$) channels were obtained by using theoretical calculations~\cite{Arvieux,VanOers,Huttel,Kievsky,Black} and a vast collection of scattering data~\cite{PhysRevLett.71.3762,PhysRev.72.662,BRUNE199813,HUTTEL1983435,CLEGG1995200}. The positive sign of the \pd scattering parameters reported in Table~\ref{tab:ScatteringLength_dp_kd} corresponds to a repulsive interaction for the quartet state ($S = 3/2$), but for the doublet state ($S = 1/2$), the $^{3}$He bound state emerges, and the standard effective range expansion has to be modified (see  Eq.~2 in Ref.~\cite{Kievsky}).

\begin{table}[h]
\begin{center}
    \caption{Scattering lengths $a_0$ and effective ranges $d_0$ for the \pd and \kd $s$-wave states. For the \pd, the two spin states (doublet and quartet) are reported. For the meson--baryon system, the negative values of scattering length refer to a repulsive interaction. For the baryon--baryon system, negative and positive values of $a_0$ refer to attractive and repulsive interactions (for cases where the potential does
not support two-body bound states), respectively.}
{\renewcommand{\arraystretch}{1.35}
    \begin{tabular}{|c|cc|cc|cc|cc|c|}
    \hline
     \multirow{2}{*}{System} & \multicolumn{2}{c|}{Spin averaged} &\multicolumn{2}{c|}{$S = 1/2$ } & \multicolumn{2}{c|}{$S = 3/2$ } & \multirow{2}{*}{References} &\\
       &$a_{0}$(fm) & $d_{0}$(fm) &  $a_{0}$(fm) & $d_{0}$(fm) &  $a_{0}$(fm) & $d_{0}$(fm) &&\\
    \hline
  \multirow{2}{*}{\kd} & $-0.470$ &  $1.75$ &--- &---& --- &---&ER~\cite{Takaki:2009ei}   & \\
    & $-0.540$  & $0.0$ & ---&--- &---  & ---&FCA~\cite{Aoki:2018wug,Kamalov:2000iy}  & \\
    \hline
 \multirow{5}{*}{\pd}
    & & &
    $2.73^{+0.10}_{-0.10}$& 2.27$^{+0.12}_{-0.12}$ & $11.88_{+0.40}^{-0.10}$ & 2.63$^{+0.01}_{-0.02}$  & Arvieux~\cite{Arvieux}& \\
    & & & $1.30^{+0.20}_{-0.20}$ & --- & $11.40^{+1.80}_{-1.20}$ & 2.05$^{+0.25}_{-0.25}$ & VanOers~\cite{VanOers}&\\

    & & &
    $4.0$ & --- & $11.1$ & --- & Huttel~\cite{Huttel}& \\
    & & &
    $0.024$& --- & $13.8$ & --- &Kievsky~\cite{Kievsky}& \\
    & & &
    $-0.13^{+0.04}_{-0.04}$ & --- & $14.70_{-2.30}^{+2.30}$ & --- &Black~\cite{Black}& \\
    \hline
    \end{tabular}}
    \label{tab:ScatteringLength_dp_kd}
\end{center}
\end{table}

An alternative method to test the accuracy of the relative wave function in a two-hadron system is the measurement of
the correlation function among the pairs of interest produced in hadron--hadron collisions~\cite{ALICE:pOmega}.
The theoretical correlation function can be expressed~\cite{Pratt:1986cc,Lisa:2005dd} as $C(\ks)\,=\,\int d^3\rs S(\rs)|\psi(\vec{\ks},\vec{\rs})|^2$,
where $S(\rs)$ is the distribution of the distance \rs between the emitted particles in a hadron--hadron collision defining the particle source,
$\psi(\vec{\ks},\vec{\rs})$ represents the wave function of the relative motion for the pair of interest, and $\ks$ is the reduced relative momentum of the pair ($\ks\,=\,|\vec{p^*_2}-\vec{p^*_1}|/2$).
The asterisk indicates that the quantities are evaluated in the pair rest frame, where $\vec{p^*_1} = - \vec{p^*_2}$.
The \LedLu ~(LL) formalism~\cite{Lednicky:1981su,Lednicky:2005tb} provides a simplified analytical treatment of the
wave function (see Appendix~\ref{sec:Methods} for details) that can relate the correlation function to its asymptotic behavior where the core nuclear strong interaction is not considered.
The analytical formula for the correlation function is obtained assuming a Gaussian source distribution in  \rs, and a single set of scattering parameters. The formula is averaged over spin and isospin and considers only $s$-waves in the scattering process.

Experimentally, the correlation function is defined as $C(\ks)\,=\,\xi(\ks)\otimes\frac{N_{\mathrm{same}}(\ks)}{N_{\mathrm{mixed}}(\ks)}$, where $\xi(\ks)$ denotes the corrections for experimental effects (see Appendix~\ref{sec:Methods} for details),
$N_{\mathrm{same}}(\ks)$ is the number of detected particle pairs in a given \ks interval obtained by combining particles produced in the same collision (event), which constitute a sample of correlated pairs, and $N_{\mathrm{mixed}}(\ks)$ is the number of uncorrelated pairs in the same \ks interval, obtained by combining particles produced in different collisions
(mixed events).
In this work, the interest resides in studying the final-state interaction for \kd and \pd pairs produced in pp collisions at $\sqrt{s}=13$~TeV.
The analyzed data set is collected using an online trigger to select high-multiplicity pp collisions to enhance the pair sample size.
Kaon (\K), antikaon (\Kbar), proton (p), antiproton (\pbar), deuteron (d), and antideuteron (\dbar) tracks are reconstructed with the ALICE detector, and their momentum in the laboratory frame \plab~ is measured in the range \plab~ $\in$ $[0.2,4.1]$~\GeVc.
The particle identification is carried out using measurements of the specific energy loss in the Time Projection Chamber (TPC) and time-of-flight (TOF) detector, resulting in samples of \K (\Kbar), p (\pbar), d (\dbar) with a purity of \purityK\%~(\purityAK\%), \purityp\%~(\purityAp\%), and \purityd\%~(\purityAd\%) respectively, as estimated via Monte Carlo simulations.
Details on the experimental methods and the evaluation of the systematic uncertainties are described in Appendix~\ref{sec:Methods}.
Once the kaons, protons, and deuterons (and charge conjugates) are selected and their three-momenta  measured, the correlation functions can be built.
Since it is assumed that the same interaction governs hadron--hadron and antihadron--antihadron pairs~\cite{ALICE:Run1}, in the following, the sum of particles and antiparticles is considered ($\kd \equiv \kd\oplus \AkAd$ and  $\pd  \equiv \pd\oplus\ApAd$).

\begin{figure*}[ht]
\centering
\subfigure{\centering
  \def\big{\includegraphics[width=0.48\linewidth]{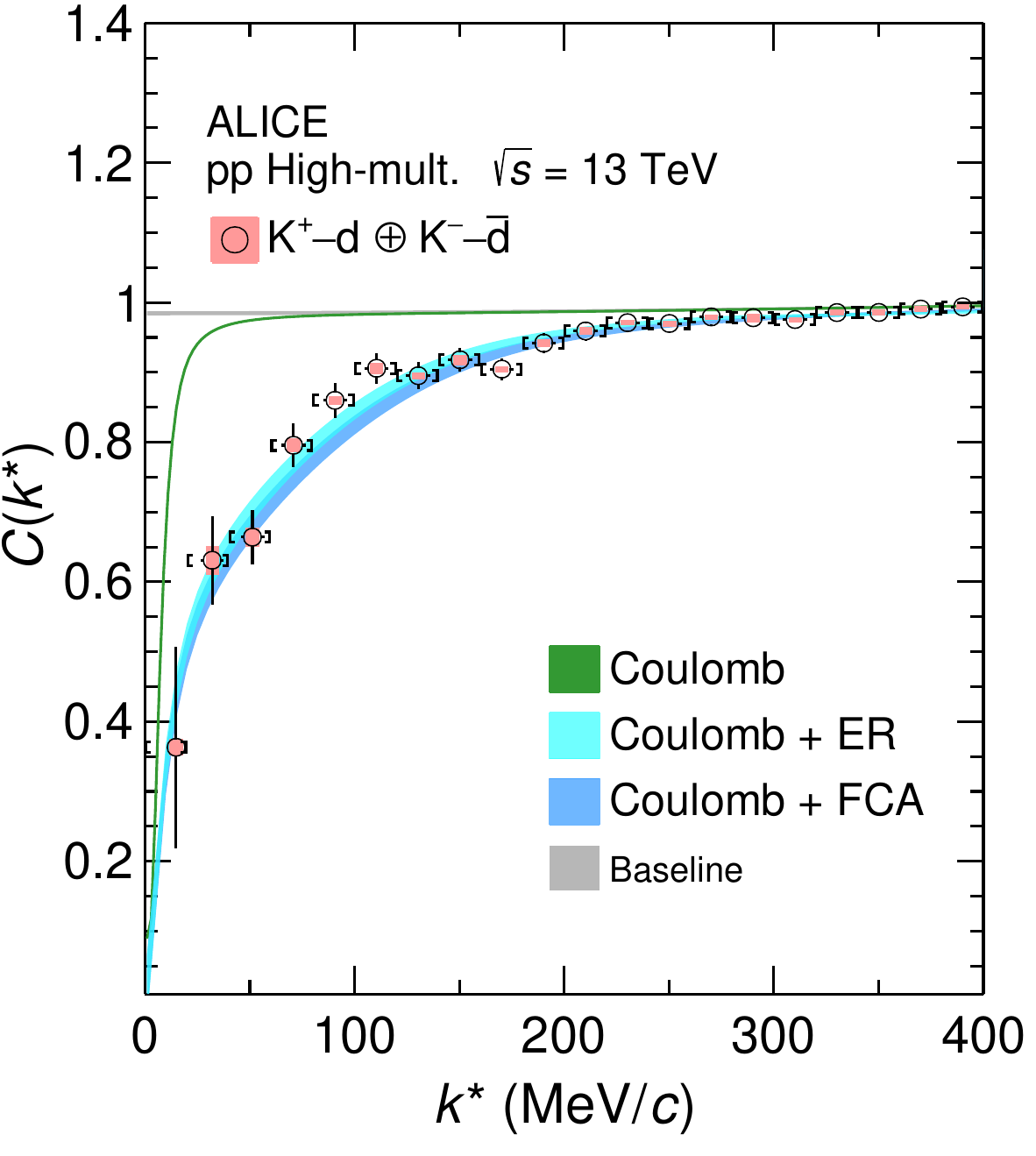}}
\def\little{\includegraphics[height=1.2cm]{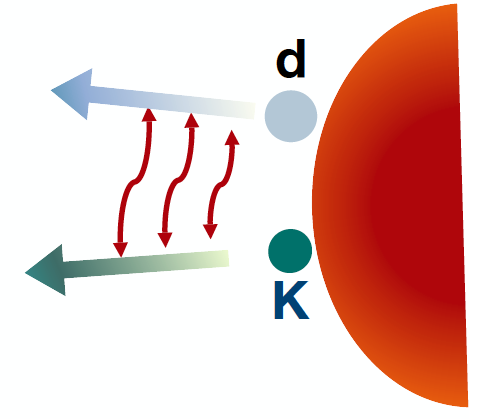}}
\stackinset{r}{30pt}{t}{100pt}{\little}{\big}}
\subfigure{\centering
  \def\big{\includegraphics[width=0.48\linewidth]{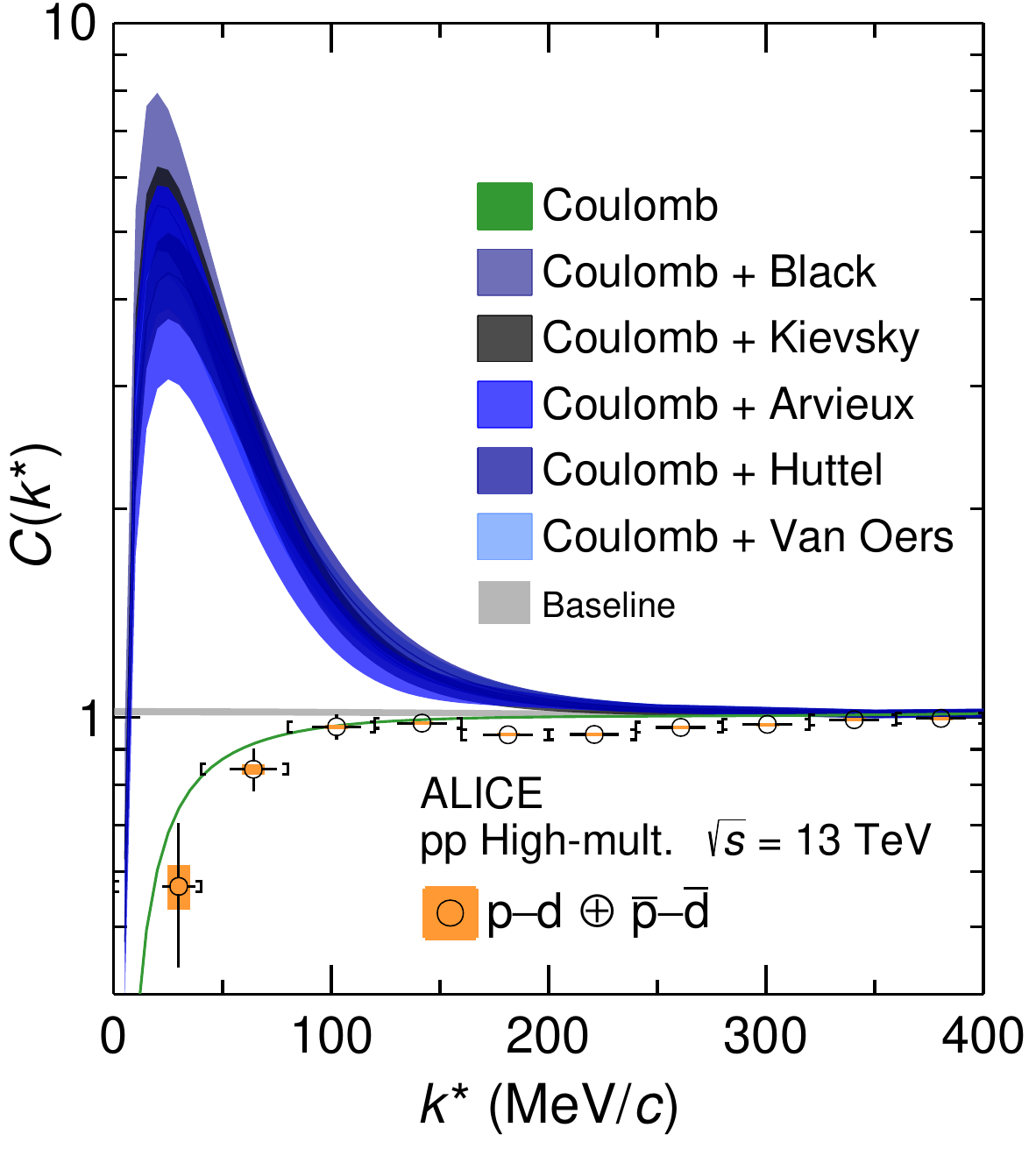}}
\def\little{\includegraphics[height=1.2cm]{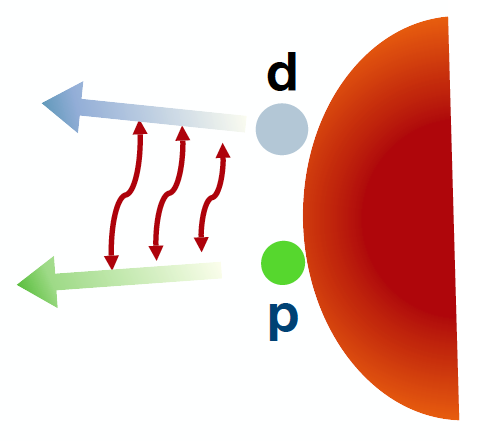}}
\stackinset{r}{20pt}{t}{14pt}{\little}{\big}}
\hfill
\caption{Measured \kd (left) and \pd (right) correlation functions. The data are shown by the black symbols, the bars and the colored boxes represent the statistical and systematic uncertainties, respectively. The square brackets show the bin width of the measurement, while the horizontal black lines represent the statistical uncertainty in the determination of the mean $\ks$ for each bin. Data are compared with theoretical correlation functions, shown by colored bands, obtained using the LL approximation. The bandwidths represent the uncertainties in the determination of the radius and the residual contributions. See text for details.}
 \label{fig:expKdPd}
\end{figure*}

Figure~\ref{fig:expKdPd} shows the \kd (left panel) and \pd (right panel) correlation functions as a function of \ks measured in pp collisions at $\sqrt{s}=13$~TeV with ALICE. Both correlation functions are below unity for values of \ks smaller than 200 \MeVc, indicating an overall repulsive interaction.
The measured correlation functions are compared to calculations performed using the LL approximation considering either only the Coulomb interaction or, in addition, the strong interaction part determined by the scattering parameters reported in Table~\ref{tab:ScatteringLength_dp_kd} for the \kd and \pd systems.
In order to compare the experimental data to the LL calculations, the source term included in the formula of the correlation function and the feed-down corrections due to particle decays and residual background contributions are needed to match the experimental measurements.

The source term has been approximated as a Gaussian distribution, whose width, defining the source size, needs to be evaluated.
The values of the \pd and \kd source sizes have been obtained from the results of independent analyses of \pp, \kp, and \pipi
correlations, which have demonstrated the existence of a universal source for any hadron--hadron pair in pp collisions at the LHC~\cite{Acharya:2020dfb, ALICE:2023sjd} and have indicated that the source size decreases with an increasing value of the pair transverse mass $m_\mathrm{T}$ (see Appendix~\ref{sec:Methods}).
In addition, further modifications of the source distribution due to strong decays of short-lived resonances decaying into protons and kaons have been taken into account.
For \pd pairs, an effective source size of $r_\text{eff}^{\small{\pd}} = $~\rpdeff fm has been obtained. For the \kd pairs, the contribution of broad resonances with very different decay times has been taken into account, and the effective source has resulted in $r_\text{eff}^{\small{\kd}} =$~\rKdeff fm.
Note that such small source sizes imply that the most probable distance between particles is around 2 fm~\cite{Mihaylov:2018rva}.

The calculated correlation functions using the LL method, shown by the colored bands in Fig.~\ref{fig:expKdPd}, have been corrected for feed-down from weak and strong decays and residual background contributions, using the data-driven methods described in Refs.~\cite{ALICE:Run1,ALICE:pLambda}, respectively. The relative contributions from primarily produced \kd and \pd pairs to the corresponding inclusive sample have been obtained experimentally and are 0.92 and 0.82, respectively.
The residual background contribution is shown as the grey-colored band close to unity. The width of the theoretical bands stems from the uncertainty propagation of the experimental determination of the source size, feed-down, and residual background. Additional information on the source size and the different contributions to the correlation function is provided in Appendix~\ref{sec:Methods}.

The measured \kd correlation function is lower than the Coulomb-only prediction showing the presence of a repulsive strong interaction. The calculations with the LL approximation using both sets of scattering parameters (ER and FCA) provide an excellent description of the experimental correlation function within uncertainties.
Indeed, since both the Coulomb and the strong \kd interactions are repulsive at any distance and there is no feature of the strong interaction that manifests only at very short distances, an asymptotic description is sufficient to reproduce the data.
Additionally, kaons are bosons and the proton and neutron that constitute the deuteron are fermions; hence, the LL approximation of point-like and distinguishable particles is pertinent, and the properties of the deuteron are mapped in the \kd scattering parameters. Hence, given a known interaction and the precise ALICE data, from the expression $C(\ks)\,=\,\int d^3\rs S(\rs)|\psi(\vec{\ks},\vec{\rs})|^2$, one can estimate the source function, even when deuterons are at play.
The agreement between the model and the data, obtained for the small radius ($r_\text{eff}^{\small{\kd}} =$ \rKdeff fm) that follows the same $m_\mathrm{T}$ scaling as all other hadron pairs, shows that deuterons are produced at small distances with respect to other hadrons in pp collisions at the LHC, and this result provides an excellent reference for the source term of correlations involving deuterons.

Instead, a huge discrepancy is observed when comparing the \pd data to analogous calculations which consider protons and deuterons as distinguishable point-like particles and employing the small source size obtained from the $m_\mathrm{T}$ scaling. This discrepancy can be seen by comparing the measured \pd correlation in the right panel of Fig.~\ref{fig:expKdPd} to the five different blue shaded curves which are obtained using the five sets of scattering lengths reported in Table~\ref{tab:ScatteringLength_dp_kd}.

The limitations of the LL approximation in describing in detail the \pd interaction are several. The existence of the $^3{\rm He}$ bound state introduces a particular short-range behavior in the doublet state due to orthogonality requirements. Moreover, the spin structure of the quartet state is completely symmetric implying as well a specific short-range behavior of the corresponding spatial part to fulfill the requirements of the Pauli principle. In addition, the correct antisymmetrization of the wave function is not considered in the LL approximation, and such short-range features of the interactions are not taken into account.
On the other hand, the correlation function obtained with the Coulomb-only assumption (green curve in the right panel of Fig.~\ref{fig:expKdPd}) catches the correct amplitude of the experimental \pd correlation function despite the sizable scattering parameters reported in Table~\ref{tab:ScatteringLength_dp_kd}. This apparent mismatch is due to the fact that in this case, the Coulomb repulsion between the proton and the deuteron reduces the impact of the short-range part of the wave function in the correlation function, and the three-fermion system is not correctly treated within the LL approximation even if only the Coulomb interaction is considered.

%% file: Content/4_PisaProj.tex
\section*{Correlation function of a three--body system}

In order to correctly describe the three-body system p--(pn) the microscopic \pd wave function must be employed in the calculation of the \pd correlation function. The latter has been obtained by projecting the \pd wave functions on the initial three-nucleon state created after the pp collisions. The relevant source term in this calculation depends on an effective nucleon--nucleon source radius $r_\text{eff}^\text{NN}$~\cite{Mrowczynski:2020ugu,Mrowczynski:2021bzy}, since the single nucleons are the relevant degrees of freedom. Details of the calculation are presented in Appendix~\ref{sec:Methods} and in Ref.~\cite{Viviani:2023kxw}.

\begin{figure*}[hbt]
\subfigure{\centering
  \def\big{\includegraphics[width=0.49\linewidth]{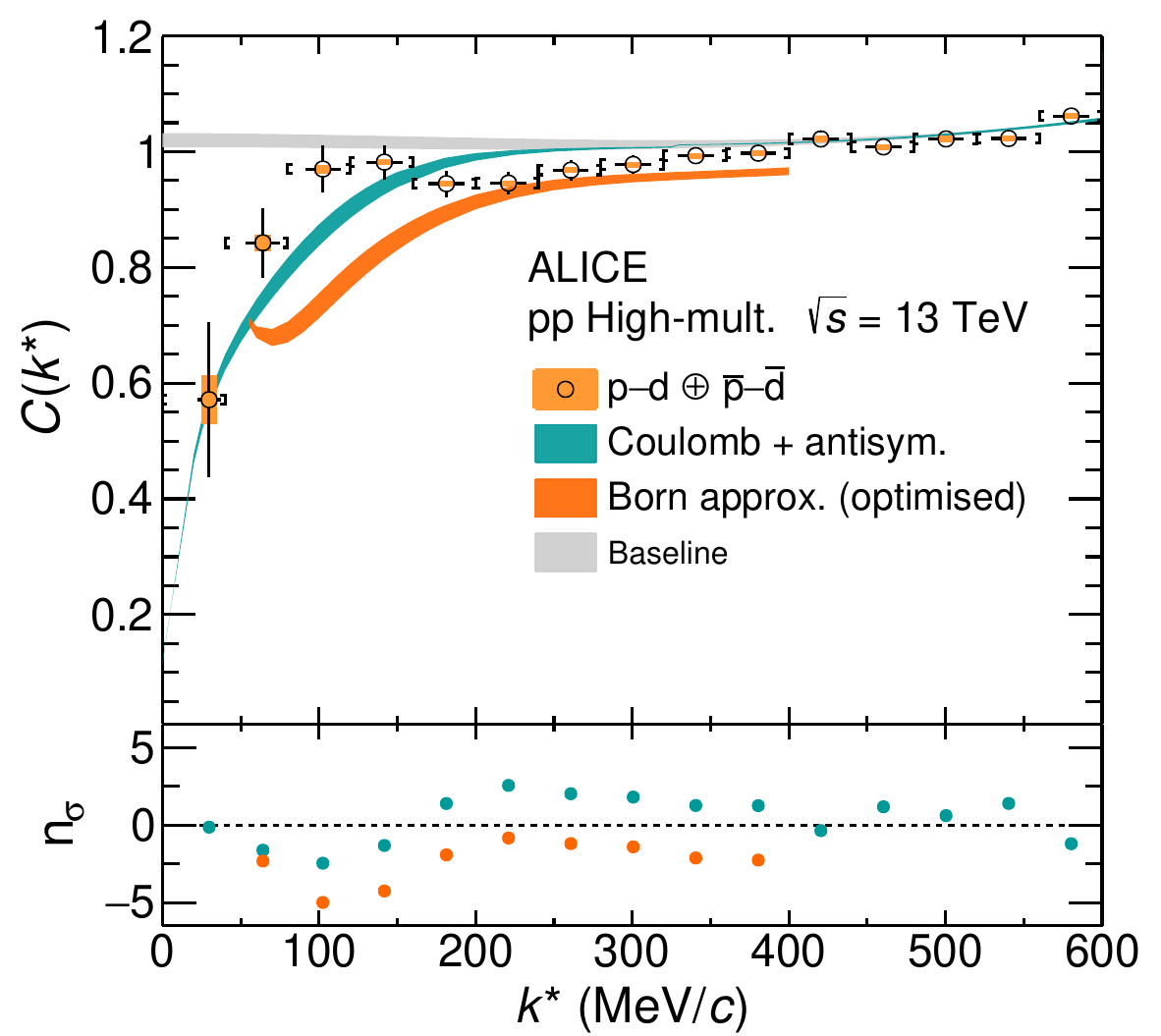}}
\def\little{\includegraphics[height=1.2cm]{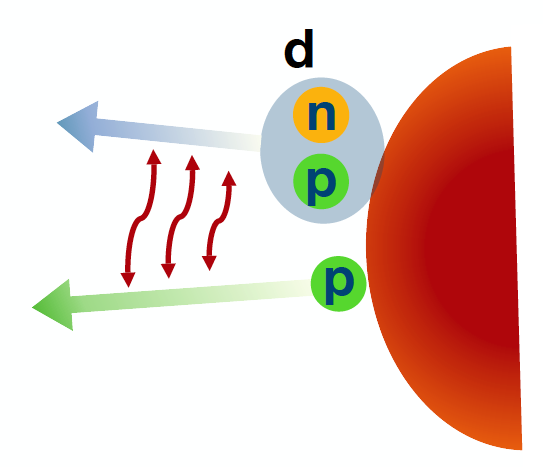}}
\stackinset{r}{135pt}{t}{94pt}{\little}{\big}}
\subfigure{\centering
  \def\big{\includegraphics[width=0.49\linewidth]{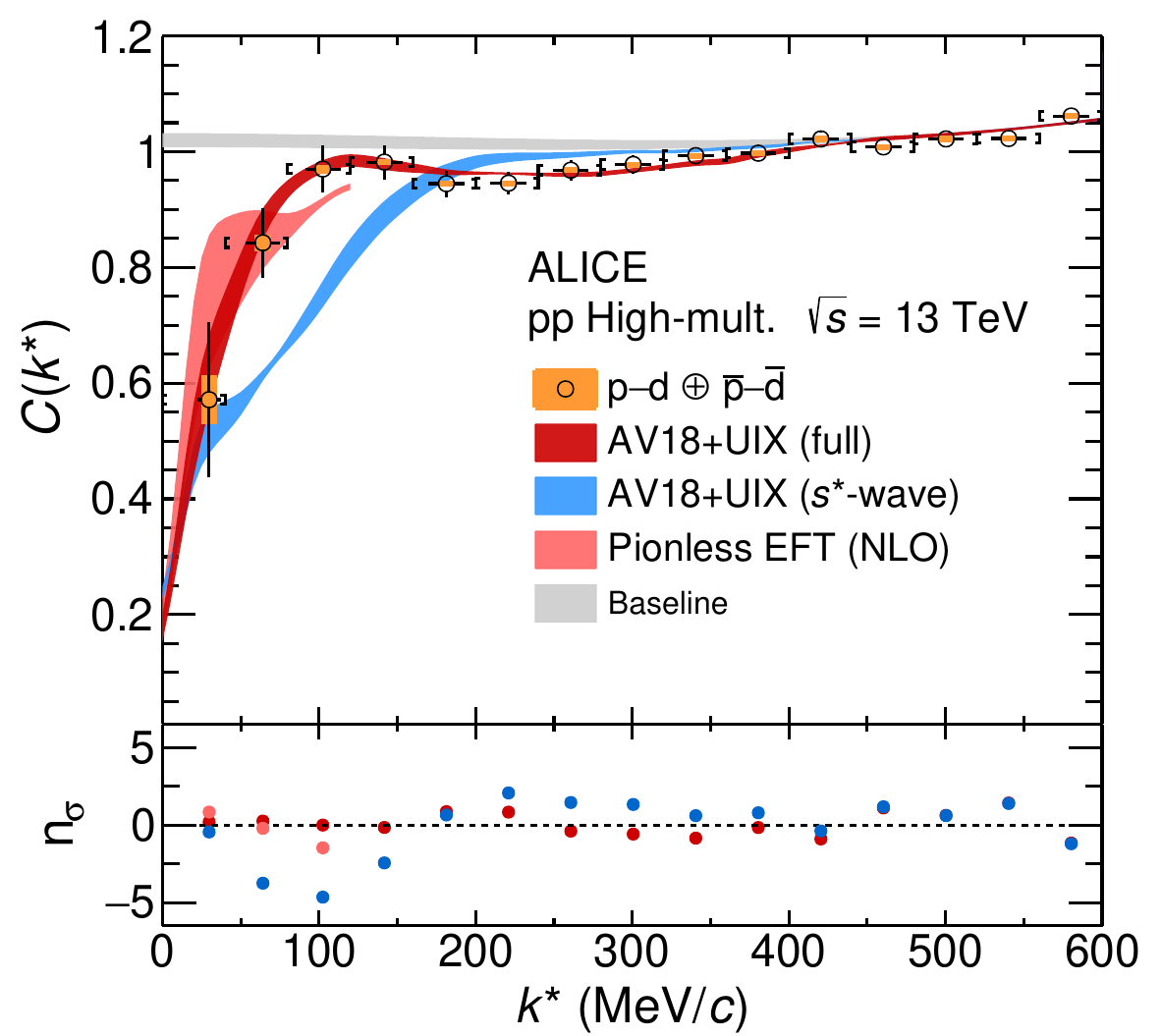}}
\def\little{\includegraphics[height=1.2cm]{Figures/Sketch_three_body_pd.png}}
\stackinset{r}{135pt}{t}{84pt}{\little}{\big}}
\hfill
\caption{Measured \pd correlation function plotted as a function of the \pd relative momentum $k^*$ alongside theoretical calculations. The experimental data are represented by circular symbols. The black vertical bars and orange boxes correspond to the statistical and systematic uncertainties, respectively. The square brackets indicate the measurement bin width and the horizontal black lines represent the statistical uncertainty in the determination of the mean $\ks$ for each bin.. The non-femtoscopic background contributions are represented by the gray band of the cubic baseline. Left panel: the orange and turquoise bands depict calculations obtained using an optimized Born approximation and Coulomb + antisymmetrization of the three-particle wave function, respectively. Right panel: the dark red band represents a fit of the modeled correlation calculated considering \pd as a three-body system with all relevant partial waves (see text). The blue-colored band corresponds to a calculation that includes only $J^\pi = \frac{1}{2}^+,\frac{3}{2}^+$ partial waves relative to the \pd system, which are dominated by the $s$-wave contributions~\cite{Viviani:2023kxw} and thus labeled with $s^*$. The light red band represents a calculation of the correlation function using Pionless EFT at NLO (see text). All calculations are multiplied by the cubic baseline, and the bandwidths of all calculations account for uncertainties in the determination of the radius and residual contributions. The lower panels present the difference between the measured and calculated correlation function, expressed as the number of standard deviations, $\text{n}_\sigma$, taking into account the statistical uncertainties of the data and the model uncertainties.}
 \label{fig:TheoryPD}
\end{figure*}

Three different versions of the \pd wave function have been investigated to study the microscopic behavior of the \pd system. First, a Born approximation of the \pd wave function that contains the correct antisymmetrization for the p--(pn) system but where the short-range contribution of the wave function has been omitted. Hence, this calculation accounts only for the asymptotic part of the wave function similar to the LL model but considers the microscopic structure of the \pd system. 
 
The left panel of Fig.~\ref{fig:TheoryPD} shows how the Born approximation compares to the experimental data, including $\text{n}_\sigma$ values that quantify the data-model deviation. It can be seen that this calculation is not sufficiently accurate to reproduce the data, although the antisymmetrization is correctly accounted for. For \ks below 60~\MeVc, this calculation predicts non-physical values and therefore they are excluded from the figure.
In the same panel, the comparison to the full-fledged Coulomb-only calculation considering the dynamics of three nucleons in the \pd system is shown as well, indicating also a clear disagreement with the data.

The second wave function that has been tested has been obtained employing the Hyperspherical Harmonics (HH) method~\cite{Kievsky:2001fq}. It accounts for all the relevant two- and three-body interactions at work in the p--(pn) system for the short and the asymptotic range, it accurately describes the three-body dynamics, and it is calculated using \pd scattering observables~\cite{Kievsky:2001fq,PhysRevC.69.014002,Deltuva:2005xa}. 
The nuclear interaction includes the AV18 two-nucleon (NN)~\cite{Wiringa:1994wb} plus the Urbana IX (UIX) three-nucleon (NNN)~\cite{Pudliner:1997ck} and the Coulomb potentials.
The blue curve in the right panel of Fig.~\ref{fig:TheoryPD} has been obtained by including the NN and NNN interactions only in the $J^\pi = \frac{1}{2}^+,\frac{3}{2}^+$ partial waves relative to the \pd system, which are dominated by the $s$-wave contributions~\cite{Viviani:2023kxw}. The $\text{n}_\sigma$ distribution in the lower panel shows that this calculation describes the data moderately well but fails in the small relative momentum part.
The agreement improves when more partial waves up to $J^\pi=\frac{7}{2}^-$ are included in the calculation, where the $p$-waves contribute predominantly, as it is shown by the
red curve of the right panel of Fig.~\ref{fig:TheoryPD}. The correlation function from the full calculation is multiplied by a baseline (gray curve in Fig.~\ref{fig:TheoryPD}) that describes the residual background. The parameters of the baseline are obtained by fitting the data and driven by the large \ks region, see Appendix~\ref{sec:Methods} for details. The same background is used for the comparison of the other calculations, and the curves' widths represent the propagated uncertainty of the source parameter and baseline. The full calculation describes the experimental data very accurately, as indicated by the $\text{n}_\sigma$ values for the red band remaining consistently close to or below 1 across the entire range of \ks.

As an additional check, the light red band in the right panel of Fig.~\ref{fig:TheoryPD} shows the \pd correlation function calculated in pionless effective field theory~\cite{Hammer:2019poc} (Pionless EFT) calculation at next-to-leading order (NLO). The nuclear interaction within this approach is much simpler than the AV18+UIX potential~\cite{Viviani:2023kxw}; the NN interaction is determined by only the $s$-wave NN scattering lengths and effective ranges up to NLO, while the NNN interaction is fixed by either the \isotope[3]{H} binding energy or the n--d $s$-wave scattering length.
For the Pionless EFT calculation, the additional uncertainty from truncating the EFT expansion at NLO can be estimated as 10\%. Taking this into account, in the regime where the theory is applicable ($k^*$ below the pion mass of approximately 140~\MeVc), the Pionless EFT results are largely compatible with the HH calculation using the AV18+UIX force.

\begin{figure}
    \centering
    \includegraphics[width=0.5\linewidth]{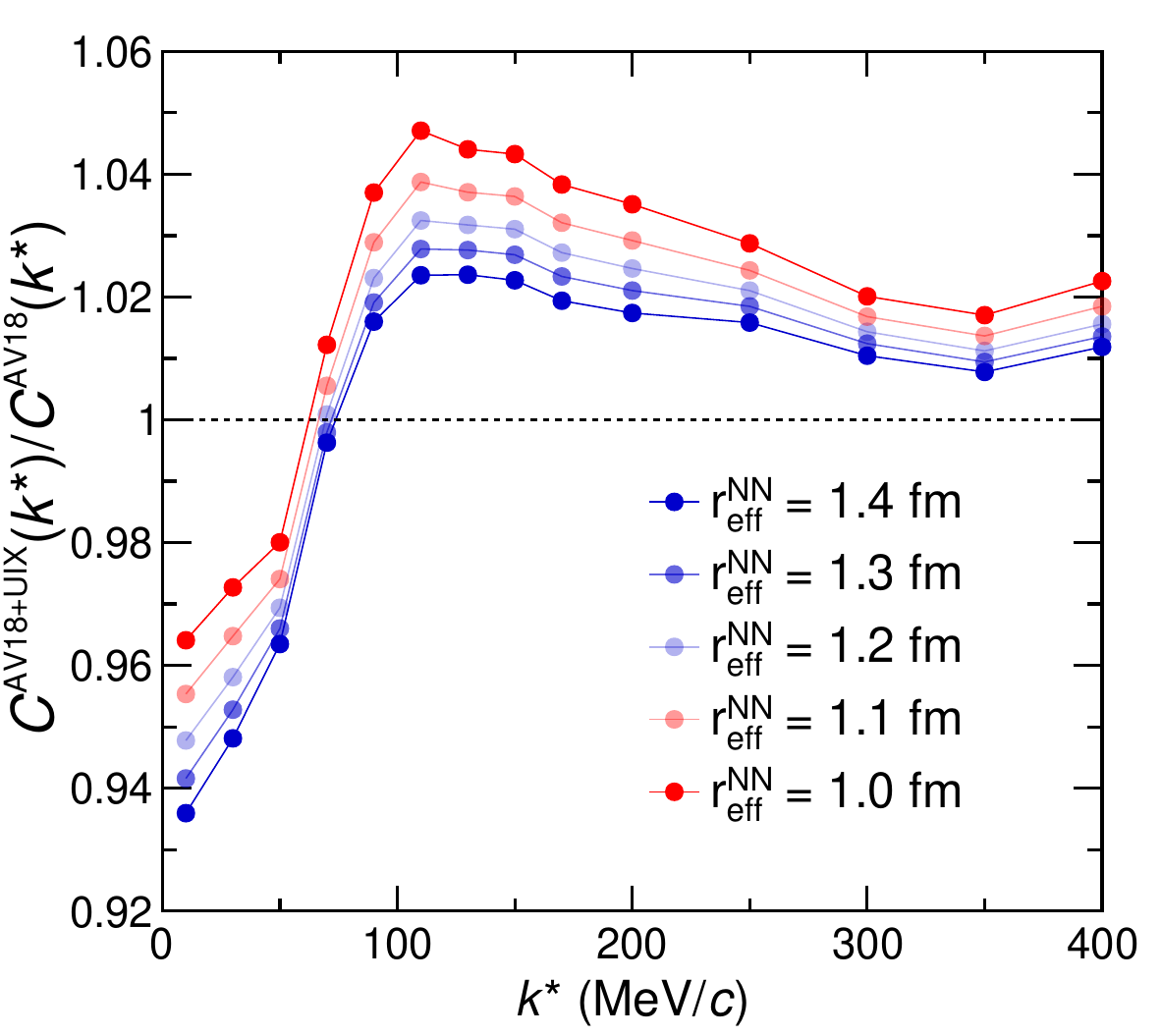}
    \caption{Ratio of theoretical \pd correlation functions obtained with full three-body calculations (AV18 for NN + UIX for NNN interaction) to those obtained using an AV18 for NN interaction only for different values of the two-nucleon effective source size $r_\text{eff}^{\text{NN}}$.}
    \label{fig:Fig3Nvs2NRatio}
\end{figure}

The fact that the experimental \pd  correlation function can be described only by a full-fledged calculation of a three-body system that considers the single nucleons as active degrees of freedom and that the shape of the correlation function is also sensitive to the inclusion of different partial waves in the interaction demonstrates for the first time that deuteron--proton correlations measured in pp collisions at the LHC access the three-nucleon system at short distances. Indeed, past measurements of correlations involving deuterons~\cite{Chitwood:1985zz,Pochodzalla:1986ova,Pochodzalla:1987zz,Wosinskapd} have been interpreted by means of the LL calculation combined with a value of the source radius larger than the one found for other hadron--hadron pairs in the same colliding system. This misinterpretation of the data was caused by the lack of correct microscopic calculations of the \pd system. It has to be observed that the current formulation of the correlation function does not guarantee invariance under unitarity transformations for distances below 1 fm, and hence the sensitivity of the observable is limited to the range $r^*>1$ fm. For the specific data-set analyzed in this work, the distances obtained for the pair of interest are mostly above 1.5 fm.

In order to test the sensitivity of the \pd correlation function to genuine three-baryon interactions,  the full-fledged \pd calculation has been carried out excluding such interactions~\cite{Viviani:2023kxw}. Figure~\ref{fig:Fig3Nvs2NRatio} shows the ratio of the calculated \pd correlation with the AV18+UIX interactions to the calculation including only the AV18 NN interactions evaluated for different values of the two-nucleon source size $r_\text{eff}^\text{NN}$.
The current precision of the data does not allow for testing such effects. However, this will be possible on the larger data samples that will be collected during the LHC Run 3 (2022--2025) with the ALICE upgraded apparatus~\cite{ALICE:2023udb}. The integrated luminosity will be increased by one order of magnitude with respect to the dataset used in the current work~\cite{YellowReport_better}; such an enhancement not only allows us to achieve precision below 1\%, similar to the case of the two-body correlation \pL~\cite{ALICE:pLambda}, but also
the differential study of the interaction at small distances by triggering on \pd pairs with large $m_\text{T}$ values. Such studies can also be extended to systems such $\Lambda/\Sigma$--d or $\Lambda^{+}_{\textrm{c}}$--d to investigate three-baryon systems in the strange and charm sectors which are otherwise inaccessible. A combined analysis of \pL and $\Lambda$--d systems will enable new experimental access to the isospin-dependent hyperon--nucleon interaction.

%% file: fa_2023-07-31_Opt_C.tex

The ALICE Collaboration would like to thank all its engineers and technicians for their invaluable contributions to the construction of the experiment and the CERN accelerator teams for the outstanding performance of the LHC complex.
The ALICE Collaboration gratefully acknowledges the resources and support provided by all Grid centres and the Worldwide LHC Computing Grid (WLCG) collaboration.
The ALICE Collaboration acknowledges the following funding agencies for their support in building and running the ALICE detector:
A. I. Alikhanyan National Science Laboratory (Yerevan Physics Institute) Foundation (ANSL), State Committee of Science and World Federation of Scientists (WFS), Armenia;
Austrian Academy of Sciences, Austrian Science Fund (FWF): [M 2467-N36] and Nationalstiftung f\"{u}r Forschung, Technologie und Entwicklung, Austria;
Ministry of Communications and High Technologies, National Nuclear Research Center, Azerbaijan;
Conselho Nacional de Desenvolvimento Cient\'{\i}fico e Tecnol\'{o}gico (CNPq), Financiadora de Estudos e Projetos (Finep), Funda\c{c}\~{a}o de Amparo \`{a} Pesquisa do Estado de S\~{a}o Paulo (FAPESP) and Universidade Federal do Rio Grande do Sul (UFRGS), Brazil;
Bulgarian Ministry of Education and Science, within the National Roadmap for Research Infrastructures 2020-2027 (object CERN), Bulgaria;
Ministry of Education of China (MOEC) , Ministry of Science \& Technology of China (MSTC) and National Natural Science Foundation of China (NSFC), China;
Ministry of Science and Education and Croatian Science Foundation, Croatia;
Centro de Aplicaciones Tecnol\'{o}gicas y Desarrollo Nuclear (CEADEN), Cubaenerg\'{\i}a, Cuba;
Ministry of Education, Youth and Sports of the Czech Republic, Czech Republic;
The Danish Council for Independent Research | Natural Sciences, the VILLUM FONDEN and Danish National Research Foundation (DNRF), Denmark;
Helsinki Institute of Physics (HIP), Finland;
Commissariat \`{a} l'Energie Atomique (CEA) and Institut National de Physique Nucl\'{e}aire et de Physique des Particules (IN2P3) and Centre National de la Recherche Scientifique (CNRS), France;
Bundesministerium f\"{u}r Bildung und Forschung (BMBF) and GSI Helmholtzzentrum f\"{u}r Schwerionenforschung GmbH, Germany;
General Secretariat for Research and Technology, Ministry of Education, Research and Religions, Greece;
National Research, Development and Innovation Office, Hungary;
Department of Atomic Energy Government of India (DAE), Department of Science and Technology, Government of India (DST), University Grants Commission, Government of India (UGC) and Council of Scientific and Industrial Research (CSIR), India;
National Research and Innovation Agency - BRIN, Indonesia;
Istituto Nazionale di Fisica Nucleare (INFN), Italy;
Japanese Ministry of Education, Culture, Sports, Science and Technology (MEXT) and Japan Society for the Promotion of Science (JSPS) KAKENHI, Japan;
Consejo Nacional de Ciencia (CONACYT) y Tecnolog\'{i}a, through Fondo de Cooperaci\'{o}n Internacional en Ciencia y Tecnolog\'{i}a (FONCICYT) and Direcci\'{o}n General de Asuntos del Personal Academico (DGAPA), Mexico;
Nederlandse Organisatie voor Wetenschappelijk Onderzoek (NWO), Netherlands;
The Research Council of Norway, Norway;
Commission on Science and Technology for Sustainable Development in the South (COMSATS), Pakistan;
Pontificia Universidad Cat\'{o}lica del Per\'{u}, Peru;
Ministry of Education and Science, National Science Centre and WUT ID-UB, Poland;
Korea Institute of Science and Technology Information and National Research Foundation of Korea (NRF), Republic of Korea;
Ministry of Education and Scientific Research, Institute of Atomic Physics, Ministry of Research and Innovation and Institute of Atomic Physics and Universitatea Nationala de Stiinta si Tehnologie Politehnica Bucuresti, Romania;
Ministry of Education, Science, Research and Sport of the Slovak Republic, Slovakia;
National Research Foundation of South Africa, South Africa;
Swedish Research Council (VR) and Knut \& Alice Wallenberg Foundation (KAW), Sweden;
European Organization for Nuclear Research, Switzerland;
Suranaree University of Technology (SUT), National Science and Technology Development Agency (NSTDA) and National Science, Research and Innovation Fund (NSRF via PMU-B B05F650021), Thailand;
Turkish Energy, Nuclear and Mineral Research Agency (TENMAK), Turkey;
National Academy of  Sciences of Ukraine, Ukraine;
Science and Technology Facilities Council (STFC), United Kingdom;
National Science Foundation of the United States of America (NSF) and United States Department of Energy, Office of Nuclear Physics (DOE NP), United States of America.
In addition, individual groups or members have received support from:
Czech Science Foundation (grant no. 23-07499S), Czech Republic;
European Research Council, Strong 2020 - Horizon 2020 (grant nos. 950692, 824093), European Union;
ICSC - Centro Nazionale di Ricerca in High Performance Computing, Big Data and Quantum Computing, European Union - NextGenerationEU;
Academy of Finland (Center of Excellence in Quark Matter) (grant nos. 346327, 346328), Finland.

%% file: Content/7_Methods.tex
\section{Methods}
\label{sec:Methods}
\subsection*{Event selection}
A data sample of inelastic pp collisions at $\sqrt{s}=13$~TeV was recorded  with ALICE~\cite{Aamodt:2008zz, Abelev:2014ffa} at the LHC.
General details on the event selection, pile-up rejection, and the primary vertex reconstruction can be found in Ref.~\cite{Acharya:2019kqn}.
A trigger that requires the total signal amplitude measured in the V0 detector~\cite{Abbas:2013taa} to exceed a certain threshold was employed to select high-multiplicity (HM) events. The V0 detector comprises two plastic scintillator arrays placed on both sides of the interaction point at pseudorapidities $\num{2.8}<\eta<\num{5.1}$ and $\num{-3.7}<\eta<\num{-1.7}$. The pseudorapidity is defined as $\eta = -\textrm{ln} \left[ \textrm{tan} \left( \frac{\theta}{2}\right) \right]$, where $\theta$ is the polar angle of the particle with respect to the proton beam axis.

At $\sqrt{s}=13$~Te\kern-.1emV\xspace, the average number of charged particles produced in HM events in the range $|\eta|<\num{0.5}$ is 30. This $\eta$ range corresponds to the region within \num{26} degrees of the transverse plane perpendicular to the beam axis. The HM events constitute only \SI{0.17}{\percent} of the pp collisions that produce at least one charged particle in the pseudorapidity range $|\eta|<\num{1.0}$. A total of \events HM events were analyzed. Additional details on the HM event selection can be found in Ref.~\cite{Acharya:2019kqn}.
\subsection*{Particle Tracking and Identification}
For the identification and momentum measurement of charged particles, the Inner Tracking System (ITS)~\cite{Aamodt:2010aa}, Time Projection Chamber (TPC)~\cite{Alme:2010ke}, and Time-Of-Flight (TOF)~\cite{Akindinov:2013tea} detectors of ALICE are employed. All three detectors are located inside a uniform magnetic field generated by the L3 solenoid (\SI{0.5}{T}), leading to a bending of the trajectories of charged particles. The measurement of the curvature is used to reconstruct the particle momenta. Typical transverse-momentum ($p_{\rm{T}}$) resolutions for kaons, protons, and deuterons vary from about \SI{2}{\percent} for tracks with $\pt = \SI{10}{}$~GeV/$c$ to below \SI{1}{\percent} for $\pt<\SI{1}{}$~GeV/$c$. The particle identity is determined by the energy lost per unit of track length inside the TPC detector and by the particle velocity measured in the TOF detector. For additional experimental details, see Ref.~\cite{Abelev:2014ffa}. Basic details on the selection criteria of the proton, kaon, and deuteron tracks used in this work can be found in Ref.~\cite{ALICE:2023gxp}.

Kaons are identified via the measurement of the specific energy loss in TPC within a momentum range $\plab\in [0.2,2.0]~\GeVc$. This information is combined with the time-of-flight measurement for momentum $\plab > 0.4$ \GeVc. In the kaon sample, the contamination from electrons produced via photon conversion in the detector material is removed by excluding kaon candidates in the momentum range $0.5 < \plab < 0.65$ \GeVc. The selected candidates that simultaneously fulfill the selection criteria of the pion (using TPC and TOF information) or proton (using TPC information) are also excluded from the sample.

Protons are selected within a transverse-momentum range of $0.5<\pt< 4.05$~\GeVc.
They are identified by requiring TPC information for candidate tracks with momentum $p<\SI{0.75}{}$~\GeVc, while TPC and TOF information are both required for candidates with $p>\SI{0.75}{}$~\GeVc.  
The identification of deuterons is performed with the TPC in the momentum range $0.4 < \plab < 1.4$ \GeVc.
For the analysis of \kd pairs, the deuteron acceptance is extended by using TOF identification in the region $1.4 < \plab< 2.3~\GeVc$.

The selection of kaons, protons, and deuterons constitute the main source of systematic uncertainties associated with the measured correlation function. All particle selection criteria are varied with respect to their default values. In order to account for the effect of possible correlations, the analysis of \kd (\pd) pairs is repeated 40 (44) times using random combinations of such selection criteria. The associated systematic uncertainty of the measured correlation function for each \ks point is given by the root-mean-square of all 40 (44) resulting correlation functions. The total systematic uncertainties are maximal at low \ks, reaching a value of \SI{3}{\percent} and \SI{7}{\percent} for \kd and \pd, correspondingly. 

\subsection*{Characterization of the source of particles}
The distribution $S(\rs)$ of the distance \rs at which particles are emitted, is described by a Gaussian function whose width characterizes the source size. In Ref.~\cite{Acharya:2020dfb}, the baryon--baryon source sizes are determined as a function of the transverse mass of the baryon--baryon pair, $\smash{\mt = \left(\kt^{2} + m^{2}\right)^{1/2}}$, where $m$ is the average mass, and $\smash[b]{\kt =\,\,\mid \bm{p}\bf{_{\textrm{T},~\textrm{1}}} + \bm{p}\bf{_{\textrm{T},~\textrm{2}}}\mid}$/2 is the transverse momentum of the pair. For \pd and \kd pairs, it is assumed that the common source for all baryons still holds for any hadron--hadron pair~\cite{Acharya:2020dfb} .
The measured $\langle m_{\mathrm{T}}\rangle$ for \pd and \kd pairs are \SI{1.64}{\GeVc} and \SI{1.50}{\GeVc}, respectively, implying source sizes of  $r_\text{core}^{\small\kd} =~$\rKdcore fm and $r_\text{core}^{\small\pd} = $ \rpdcore fm, where $r_{\textrm{core}}$ denotes the width of the Gaussian distribution defining the source before taking into account the effect produced by short-lived resonances.

In pp collisions at \onethree, about $2/3$ of the protons and $1/3$ of the kaons originate from the decay of short-lived resonances with a lifetime $(c\tau)$ of a few fm. Table~\ref{tab:TabReso} shows the major relative contributions to the proton and kaon samples by several resonances, according to statistical hadronization~\cite{VOVCHENKO2019295}. The effect of such resonances on the source size is evaluated by folding the Gaussian source with an exponential distribution containing the resonance decay constant, following Ref.~\cite{Acharya:2020dfb}.
The resulting source distribution for \pd can be parameterized with an effective Gaussian source radius equal to $r_\text{eff}^{\small{\pd}} =$~\rpdeff\si{\fm}. For \kd pairs, the resulting source distribution is described using a sum of two Gaussian functions with radii $1.10 \pm 0.04$~\si{\fm} and $2.14^{+0.03}_{-0.07}$~\si{\fm} with weights of $0.76$ and $0.24$, respectively. The effective source size quoted in the text $r_\text{eff}^{\small{\kd}} =$~\rKdeff~\si{\fm} results from the weighted sum of the two Gaussians. 
Note that a description with a single Gaussian (1G) for the \kd source distribution, including the effects of resonances is not satisfactory, but it is nevertheless illustrative to compare the resulting radius of $r_\text{eff1G}^{\small{\kd}} =$~\rKdeffoneg~\si{\fm} with the \pd radius $r_\text{eff}^{\small{\pd}} =$~\rpdeff\si{\fm}.

In the case of the three-body calculation for the \pd system, the formalism requires as input the two-particle source size for each pairwise source of the p--(pn) system. For nucleon pairs, an average core source size of $r_\text{core}^\text{NN} =$~\rNNcore fm is estimated using the measured $\langle m_{\mathrm{T}}\rangle$ for \pd and the measured $\langle p_{\mathrm{T}}\rangle$ of deuteron and protons below $k^*<400$~\MeVc. The source size  $r_\text{core}^\text{NN}$ is enhanced by the strongly decaying resonances that feed to protons and neutrons. Similar to the case of \pd and \kd pairs, this effect in the source size is taken into account by folding the Gaussian source with an exponential distribution. The resulting source distribution for NN pairs can be characterized by an effective Gaussian source radius equal to $r_\text{eff}^\text{NN} = $~\rNNeff\si{\fm}. This source size is an average of distances between the p--p and p--n pairs that form the \pd system.

\begin{table}[ht]
\begin{center}
    \caption{Fractions of protons and kaons feeding from the strongly-decaying short-lived decays of resonances. The contribution of kaons from the $\phi(1020)$ decays are considered as feed-down to the correlation functions.}
    \begin{tabular}{|ccc|ccc|}
    \hline
    \multicolumn{3}{|c|}{Protons} & \multicolumn{3}{c|}{Kaons}\\
   Resonances  & $c\tau(\mathrm{fm}) $ & Fractions(\%) & Resonances  & $c\tau(\mathrm{fm}) $ & Fractions(\%)\\
    \hline
    $\Delta^{++}$ & 1.64&$21.87$&$\mathrm{K^*(892)^0}$ &  3.89 & $21$ \\
    \hline
    $\Delta^+$  &  1.64&$14.60$& $\mathrm{K^*(892)^+}$&  3.88 &$11$ \\
    \hline
    $\Delta^0$  &  1.64&$7.20$& $\mathrm{a_0(980)^+}$&  2.63 &$1$ \\
    \hline
    N$(1440)^0$ &  0.56&$0.91$ & $\mathrm{K_2^{*}(1430)^0}$&  1.81&$1$ \\
    \hline
    N$(1520)^0$  &  1.64&$1.75$& $\mathrm{K_1^{*}(1270)^0}$&  2.19 &$1$ \\
    \hline
    N$(1680)^0$  &  1.52&$1.15$& $\phi(1020)$&  46.32 &$(6)$ \\
    \hline
    N$(1535)^+$  &  1.31 &$1.02$ &$-$& $-$& $-$ \\
    \hline
    N$(1440)^+$  &  0.56 &$0.91$ & $-$ & $-$ & $-$\\
    \hline
    \end{tabular}
    \label{tab:TabReso}
\end{center}
\end{table}

\subsection*{Corrections of the Correlation Function}
The experimental correlation function, defined as $C(\ks) = \xi(\ks) \otimes \frac{N_{\mathrm{same}}(\ks)}{N_{\mathrm{mixed}}(\ks)}$, is corrected via $\xi(\ks)$ for normalization and the unfolding of momentum resolution effects. A normalization factor $\mathcal{N}$ is used to correct for the different yields in same- and mixed-event distributions. The value of the normalization factor is chosen such that the mean value of the correlation function is equal to unity. It is obtained by dividing the integral of the $N_{\mathrm{same}}(\ks)$ and $N_{\mathrm{mixed}}(\ks)$ in a region where the effect of final-state interactions is negligible (for \pd: 200 $<$ \ks $<$ 500\,\MeVc and in the case of \kd: 300 $<$ \ks $<$ 600\,\MeVc).

The shape of the measured correlation function is affected by the finite resolution in the determination of the momentum of particles. The measured same- and mixed-event distributions are unfolded for the momentum smearing using Monte Carlo (MC) simulations. The correction for the measured correlation function due to the momentum resolution effect with the MC event generator is found to be at most 3\% for \pd and 9\% for \kd at low \ks. Moreover, the resolution effects due to the merging of trajectories, which cross the detector at a distance comparable with the spatial resolution of the detector (ITS or TPC), are evaluated and found to be negligible.

While the measured \pd and \kd correlation functions are dominated by the interaction between \pd and \kd pairs consisting of primary particles produced in the collision,  they are also influenced by the effects caused by the misidentification of particles and by the contribution of secondary particles produced in weak decays. Because of the large decay times of weakly decaying particles, the final-state interaction between their decay products and the primary particle of interest is absent, thus leading to either a flat contribution or a residual signature of the interaction with the parent hadron. 
The contributions to $C(\ks)$ from the genuine correlation of primary particles, the misidentified hadrons, and the secondary particles are quantified by the so-called $\lambda$ parameters, which are computed from the purity and the primary fraction of each particle species as described in Ref.~\cite{ALICE:Run1}. The $\lambda$ parameters for the genuine correlation function of \pd and \kd are listed in \autoref{tab:LamParam}. All residual correlations not stemming from genuine primary pairs in the \pd and \kd correlation functions are assumed to be equal to unity regardless of the values of \ks, and the theoretical correlation functions shown in Figs.~\ref{fig:expKdPd} and~\ref{fig:TheoryPD} are corrected using the corresponding $\lambda$ parameters.

\begin{table}[ht]
    \centering
    \caption{$\lambda$ parameters of the genuine particle pairs for \pd and K$^+$--d.}
    \begin{tabular}{|c|c|}
     \hline
$\lambda$ parameter & Value \\
    \hline
       $\lambda_{\mathrm{pd}}$  &  79.7\% \\
     $\lambda_{\mathrm{\Bar{p}\Bar{d}}}$ & 84.1\% \\
     $\lambda_{\mathrm{K^{+}d}}$ & 90.1\% \\
     $\lambda_{\mathrm{K^{-}\Bar{d}}}$ &  94.1\%\\
     \hline
    \end{tabular}
    \label{tab:LamParam}
\end{table}

In addition to the residual correlation, the theoretical correlation function is also corrected for the rising tail effect of the experimental correlation in the region where the final-state interaction is negligible and that can be originated from effects such as energy conservation. A multiplicative cubic baseline $BL(k^*) = a + b{k^*}^2+c{k^*}^3$ is used to describe the data at large \ks.
For \kd correlations, the baseline is pre-fitted in the region of 300 $<$ \ks $<$ 1800 \MeVc. In the case of the \pd correlation function, the parameters $a$, $b$, and $c$ are obtained from a fit to the data in the region 0 $<$ \ks $<$ 700 \MeVc ~including the theoretical correlation function and the multiplicative baseline. 
The baseline is represented by the gray bands in Figs.~\ref{fig:expKdPd} and~\ref{fig:TheoryPD}.

The uncertainties on the determination of the residual contributions, the baseline, and the source size are propagated to the theoretical correlation functions and contribute, together with the intrinsic theoretical uncertainties of each model, to the width of the bands shown in Figs.~\ref{fig:expKdPd} and~\ref{fig:TheoryPD}.
The main contributions to the total uncertainty are those derived from the baseline fit and the radius determination. The former are evaluated by variations in the limits of the fit range
of $\pm 50$ \MeVc~for the upper and lower limits in the case of \kd, and $\pm 40$ \MeVc~ in the upper limit for \pd. The uncertainty of the radius is propagated from the uncertainties in the parametrization of the \mt scaling according to Ref.~\cite{Acharya:2020dfb}, and it is of 6.2\% (0.1\%) and 2\% (2\%) at \ks = 50 (100) MeV/c for \pd and \kd pairs, respectively.

\subsection*{The \LedLu Model}
The final-state interaction for two charged point-like particles has been modeled in Ref.~\cite{Lednicky:2005tb}.
The definition of the relative $s$-wave function for a system of two charged point-like distinguishable particles is given as
\begin{equation}
\label{eq:lednickyStrongCoulombWavefunction}
\psi(\vec{k}^*,\vec{r}^*) =e^{i\delta_c}\sqrt{A_{c}(\eta)}\left[e^{-i\vec{k}^*\cdot\vec{r}^*}F(-i\eta,1,i\xi)+f_{C}(k^*)\frac{\tilde{G}(\rho,\eta)}{r*}
\right]\,, 
\end{equation}
where $\eta=(k^*a_{C})^{-1}$, $a_C$ is the Bohr radius including the sign of the interaction between the pair of particles, $\xi = \rho(1+\cos(\theta^*))$, and $\rho = k^*r^*$, where $\theta^*$ is the angle between $\vec k^*$ and $\vec r^*$. The term $A_C(\eta) = 2\pi\eta \left[ \exp(2\pi\eta) - 1 \right ]^{-1} $ is a Coulomb-barrier penetration factor, also known as the Gamow factor. The Coulomb interaction (asymptotic form) in the wave function is described by the term $e^{-\iota \vec k^*\vec \cdot r^*}F(\alpha,\, 1,\, z)$ together with $\widetilde{G}(\rho,\eta)$, where $F(\alpha,\, 1,\, z) = 1 + \frac{\alpha \,z}{!2} + + \frac{\alpha (\alpha +1)\,z^2}{!2^2} +\cdots$  is a confluent hypergeometric function and $\widetilde{G}(\rho,\eta) =  \sqrt{Ac}(G_0 + iF_0)$ is a combination of
the regular $(F_0)$ and singular $(G_0)$ $s$-wave Coulomb functions. The strong nuclear interaction is computed using the Coulomb-corrected scattering amplitude $f_{\mathrm{C}}$ defined as 
\begin{equation}
\label{eq:amplitude_coulombstrong}
f_C(k^*)=\left[\frac{1}{-a_0}+\frac{d_0{k^*}^{2}}{2}-ik^*A_C(k^*)-\frac{2}{a_C} h(k^*) \right]^{-1}.
\end{equation}
The function $h(k^*)$ is written as $h(k^*)=\frac{1}{(k^*a_{C})^{2}}\sum^{\infty}_{n=1}\left[n\left(n^2+(k^*a_{C})^{-2}\right)\right]^{-1}-\gamma+\ln |k^*a_{C}|$, where the constant $\gamma=0.5772$ is the Euler constant.

\subsection*{Three-body formalism for the \pd correlation function}
In order to account for the internal structure of the deuteron, a calculation of the proton--deuteron correlation function that includes a microscopic \pd wave function  $\Psi_{m_2,m_1}(\bmx,\bmy)$ is carried out.  This wave function accounts for all the relevant two- and three-body interactions at work in the p--(pn) system for the short and asymptotic ranges. It accurately describes the three-body dynamics~\cite{Kievsky:2001fq,PhysRevC.69.014002,Deltuva:2005xa} since it is tuned on \pd scattering observables.  The variables $\bmx$ and $\bmy$ are the Jacobi vectors of the system, and they asymptotically denote the distance of the p--n system within the deuteron and the p--d distance, respectively. The indices $m_2$ and $m_1$ are quantum numbers of the spin operators for the deuteron and the proton when they are very well apart.
The three-body correlation is calculated using the formalism discussed in Refs.~\cite{Mrowczynski:2020ugu,Mrowczynski:2021bzy}. More details about the employed wave functions and the computation of the \pd correlation function can be found in Ref.~\cite{Viviani:2023kxw}. 
In general, the formalism is based on the following expressions
\begin{eqnarray}
C_{pd}(k) &= & \frac{1}{A_\mathrm{d}} {\frac{1} {6}} \sum_{m_2, m_1}
\int d^3r_1 d^3r_2 d^3r_3\; S_1(r_1) S_1(r_2) S_1(r_3) |\Psi_{m_2,m_1}|^2 \ ,\label{eq:mrowmain1} \\
A_\mathrm{d} &=& {\frac{1}{ 3}} \sum_{m_2}
\int d^3r_1 d^3r_2 \; S_1(r_1) S_1(r_2) |\phi_{m_2}|^2\ , \label{eq:mrowmain2}\\
S_1(r) &=& \frac{1} {(2\pi R_{\text{M}}^2)^{\frac{3}{2}} } e^{-r^2/2 R_{\text{M}}^2}\ , \label{eq:mrowmain3} 
\end{eqnarray} 
where $\phi_{m_2}$ is the deuteron wave function and $S_1(r)$ represents the single particle source term of radius $R_\text{M}$, while the factor 6 in the denominator of Eq.~\ref{eq:mrowmain1} takes into account the possible spin configurations. The quantity $A_\mathrm{d}$ can be related to the probability of deuteron formation in the reaction. Note that in this calculation, the accurate \pd wave function is different from the approximated form employed
in Refs.~\cite{Mrowczynski:2020ugu,Mrowczynski:2021bzy}. By considering a suitable choice of coordinates, one can demonstrate~\cite{Mrowczynski:2019yrr} that the parameter $R_M$ corresponds to the effective radius for the two-nucleon system. For the \pd system, $R_M$ is estimated considering the measured $\langle m_{\mathrm{T}}\rangle$ of the \pd pairs and the average transverse momentum $\langle p_{\mathrm{T}}\rangle$ of the deuterons and protons used for the correlation. The value $R_M=\,r_\text{eff}^\text{NN} = $\rNNeff\si{\fm} is obtained and used in the calculation.

%% file: 2023-07-31-Alice_Authorlist_2023-07-31_Opt_C.tex
\begin{flushleft} 
\small

S.~Acharya\,\orcidlink{0000-0002-9213-5329}\,$^{\rm 131}$, 
D.~Adamov\'{a}\,\orcidlink{0000-0002-0504-7428}\,$^{\rm 90}$, 
G.~Aglieri Rinella\,\orcidlink{0000-0002-9611-3696}\,$^{\rm 35}$, 
M.~Agnello\,\orcidlink{0000-0002-0760-5075}\,$^{\rm 32}$, 
N.~Agrawal\,\orcidlink{0000-0003-0348-9836}\,$^{\rm 54}$, 
Z.~Ahammed\,\orcidlink{0000-0001-5241-7412}\,$^{\rm 139}$, 
S.~Ahmad\,\orcidlink{0000-0003-0497-5705}\,$^{\rm 16}$, 
S.U.~Ahn\,\orcidlink{0000-0001-8847-489X}\,$^{\rm 75}$, 
I.~Ahuja\,\orcidlink{0000-0002-4417-1392}\,$^{\rm 40}$, 
A.~Akindinov\,\orcidlink{0000-0002-7388-3022}\,$^{\rm 146}$
% ...

\section*{Affiliation Notes}

$^{\rm I}$ Deceased\\
$^{\rm II}$ Also at: Max-Planck-Institut fur Physik, Munich, Germany\\
$^{\rm III}$ Also at: Italian National Agency for New Technologies, Energy and Sustainable Economic Development (ENEA), Bologna, Italy\\
$^{\rm IV}$ Also at: Dipartimento DET del Politecnico di Torino, Turin, Italy\\
$^{\rm V}$ Also at: Yildiz Technical University, Istanbul, T\"{u}rkiye\\
$^{\rm VI}$ Also at: Department of Applied Physics, Aligarh Muslim University, Aligarh, India\\
$^{\rm VII}$ Also at: Institute of Theoretical Physics, University of Wroclaw, Poland\\
$^{\rm VIII}$ Also at: An institution covered by a cooperation agreement with CERN\\

\section*{Collaboration Institutes}

$^{1}$ A.I. Alikhanyan National Science Laboratory (Yerevan Physics Institute) Foundation, Yerevan, Armenia\\
$^{2}$ AGH University of Krakow, Cracow, Poland\\
$^{3}$ Bogolyubov Institute for Theoretical Physics, National Academy of Sciences of Ukraine, Kiev, Ukraine\\
$^{4}$ Bose Institute, Department of Physics  and Centre for Astroparticle Physics and Space Science (CAPSS), Kolkata, India\\
$^{5}$ California Polytechnic State University, San Luis Obispo, California, United States\\
$^{6}$ Central China Normal University, Wuhan, China\\
$^{7}$ Centro de Aplicaciones Tecnol\'{o}gicas y Desarrollo Nuclear (CEADEN), Havana, Cuba\\
$^{8}$ Centro de Investigaci\'{o}n y de Estudios Avanzados (CINVESTAV), Mexico City and M\'{e}rida, Mexico\\
$^{9}$ Chicago State University, Chicago, Illinois, United States\\
$^{10}$ China Institute of Atomic Energy, Beijing, China\\
$^{11}$ China University of Geosciences, Wuhan, China\\
$^{12}$ Chungbuk National University, Cheongju, Republic of Korea\\
$^{13}$ Comenius University Bratislava, Faculty of Mathematics, Physics and Informatics, Bratislava, Slovak Republic\\
$^{14}$ COMSATS University Islamabad, Islamabad, Pakistan\\
$^{15}$ Creighton University, Omaha, Nebraska, United States\\
$^{16}$ Department of Physics, Aligarh Muslim University, Aligarh, India\\
$^{17}$ Department of Physics, Pusan National University, Pusan, Republic of Korea\\
$^{18}$ Department of Physics, Sejong University, Seoul, Republic of Korea\\
$^{19}$ Department of Physics, University of California, Berkeley, California, United States\\
$^{20}$ Department of Physics, University of Oslo, Oslo, Norway\\
$^{21}$ Department of Physics and Technology, University of Bergen, Bergen, Norway\\
$^{22}$ Department of Physics of North Carolina State University, Raleigh, North Carolina, United States\\
$^{23}$ Dipartimento di Fisica, Universit\`{a} di Pavia, Pavia, Italy\\
$^{24}$ Dipartimento di Fisica dell'Universit\`{a} and Sezione INFN, Cagliari, Italy\\
$^{25}$ Dipartimento di Fisica dell'Universit\`{a} and Sezione INFN, Pisa, Italy\\
$^{26}$ Dipartimento di Fisica dell'Universit\`{a} and Sezione INFN, Trieste, Italy\\
$^{27}$ Dipartimento di Fisica dell'Universit\`{a} and Sezione INFN, Turin, Italy\\
$^{28}$ Dipartimento di Fisica e Astronomia dell'Universit\`{a} and Sezione INFN, Bologna, Italy\\
$^{29}$ Dipartimento di Fisica e Astronomia dell'Universit\`{a} and Sezione INFN, Catania, Italy\\
$^{30}$ Dipartimento di Fisica e Astronomia dell'Universit\`{a} and Sezione INFN, Padova, Italy\\
$^{31}$ Dipartimento di Fisica `E.R.~Caianiello' dell'Universit\`{a} and Gruppo Collegato INFN, Salerno, Italy\\
$^{32}$ Dipartimento DISAT del Politecnico and Sezione INFN, Turin, Italy\\
$^{33}$ Dipartimento di Scienze MIFT, Universit\`{a} di Messina, Messina, Italy\\
$^{34}$ Dipartimento Interateneo di Fisica `M.~Merlin' and Sezione INFN, Bari, Italy\\
$^{35}$ European Organization for Nuclear Research (CERN), Geneva, Switzerland\\
$^{36}$ Faculty of Electrical Engineering, Mechanical Engineering and Naval Architecture, University of Split, Split, Croatia\\
$^{37}$ Faculty of Engineering and Science, Western Norway University of Applied Sciences, Bergen, Norway\\
$^{38}$ Faculty of Nuclear Sciences and Physical Engineering, Czech Technical University in Prague, Prague, Czech Republic\\
$^{39}$ Faculty of Physics, Sofia University, Sofia, Bulgaria\\
$^{40}$ Faculty of Science, P.J.~\v{S}af\'{a}rik University, Ko\v{s}ice, Slovak Republic\\
$^{41}$ Frankfurt Institute for Advanced Studies, Johann Wolfgang Goethe-Universit\"{a}t Frankfurt, Frankfurt, Germany\\
$^{42}$ Fudan University, Shanghai, China\\
$^{43}$ Gangneung-Wonju National University, Gangneung, Republic of Korea\\
$^{44}$ Gauhati University, Department of Physics, Guwahati, India\\
$^{45}$ Helmholtz-Institut f\"{u}r Strahlen- und Kernphysik, Rheinische Friedrich-Wilhelms-Universit\"{a}t Bonn, Bonn, Germany\\
$^{46}$ Helsinki Institute of Physics (HIP), Helsinki, Finland\\
$^{47}$ High Energy Physics Group,  Universidad Aut\'{o}noma de Puebla, Puebla, Mexico\\
$^{48}$ Horia Hulubei National Institute of Physics and Nuclear Engineering, Bucharest, Romania\\
$^{49}$ HUN-REN Wigner Research Centre for Physics, Budapest, Hungary\\
$^{50}$ Indian Institute of Technology Bombay (IIT), Mumbai, India\\
$^{51}$ Indian Institute of Technology Indore, Indore, India\\
$^{52}$ INFN, Laboratori Nazionali di Frascati, Frascati, Italy\\
$^{53}$ INFN, Sezione di Bari, Bari, Italy\\
$^{54}$ INFN, Sezione di Bologna, Bologna, Italy\\
$^{55}$ INFN, Sezione di Cagliari, Cagliari, Italy\\
$^{56}$ INFN, Sezione di Catania, Catania, Italy\\
$^{57}$ INFN, Sezione di Padova, Padova, Italy\\
$^{58}$ INFN, Sezione di Pavia, Pavia, Italy\\
$^{59}$ INFN, Sezione di Pisa, Pisa, Italy\\
$^{60}$ INFN, Sezione di Torino, Turin, Italy\\
$^{61}$ INFN, Sezione di Trieste, Trieste, Italy\\
$^{62}$ Inha University, Incheon, Republic of Korea\\
$^{63}$ Institute for Gravitational and Subatomic Physics (GRASP), Utrecht University/Nikhef, Utrecht, Netherlands\\
$^{64}$ Institute of Experimental Physics, Slovak Academy of Sciences, Ko\v{s}ice, Slovak Republic\\
$^{65}$ Institute of Physics, Homi Bhabha National Institute, Bhubaneswar, India\\
$^{66}$ Institute of Physics of the Czech Academy of Sciences, Prague, Czech Republic\\
$^{67}$ Institute of Space Science (ISS), Bucharest, Romania\\
$^{68}$ Institut f\"{u}r Kernphysik, Johann Wolfgang Goethe-Universit\"{a}t Frankfurt, Frankfurt, Germany\\
$^{69}$ Instituto de Ciencias Nucleares, Universidad Nacional Aut\'{o}noma de M\'{e}xico, Mexico City, Mexico\\
$^{70}$ Instituto de F\'{i}sica, Universidade Federal do Rio Grande do Sul (UFRGS), Porto Alegre, Brazil\\
$^{71}$ Instituto de F\'{\i}sica, Universidad Nacional Aut\'{o}noma de M\'{e}xico, Mexico City, Mexico\\
$^{72}$ iThemba LABS, National Research Foundation, Somerset West, South Africa\\
$^{73}$ Jeonbuk National University, Jeonju, Republic of Korea\\
$^{74}$ Johann-Wolfgang-Goethe Universit\"{a}t Frankfurt Institut f\"{u}r Informatik, Fachbereich Informatik und Mathematik, Frankfurt, Germany\\
$^{75}$ Korea Institute of Science and Technology Information, Daejeon, Republic of Korea\\
$^{76}$ KTO Karatay University, Konya, Turkey\\
$^{77}$ Laboratoire de Physique Subatomique et de Cosmologie, Universit\'{e} Grenoble-Alpes, CNRS-IN2P3, Grenoble, France\\
$^{78}$ Lawrence Berkeley National Laboratory, Berkeley, California, United States\\
$^{79}$ Lund University Department of Physics, Division of Particle Physics, Lund, Sweden\\
$^{80}$ Nagasaki Institute of Applied Science, Nagasaki, Japan\\
$^{81}$ Nara Women{'}s University (NWU), Nara, Japan\\
$^{82}$ National and Kapodistrian University of Athens, School of Science, Department of Physics , Athens, Greece\\
$^{83}$ National Centre for Nuclear Research, Warsaw, Poland\\
$^{84}$ National Institute of Science Education and Research, Homi Bhabha National Institute, Jatni, India\\
$^{85}$ National Nuclear Research Center, Baku, Azerbaijan\\
$^{86}$ National Research and Innovation Agency - BRIN, Jakarta, Indonesia\\
$^{87}$ Niels Bohr Institute, University of Copenhagen, Copenhagen, Denmark\\
$^{88}$ Nikhef, National institute for subatomic physics, Amsterdam, Netherlands\\
$^{89}$ Nuclear Physics Group, STFC Daresbury Laboratory, Daresbury, United Kingdom\\
$^{90}$ Nuclear Physics Institute of the Czech Academy of Sciences, Husinec-\v{R}e\v{z}, Czech Republic\\
$^{91}$ Oak Ridge National Laboratory, Oak Ridge, Tennessee, United States\\
$^{92}$ Ohio State University, Columbus, Ohio, United States\\
$^{93}$ Physics department, Faculty of science, University of Zagreb, Zagreb, Croatia\\
$^{94}$ Physics Department, Panjab University, Chandigarh, India\\
$^{95}$ Physics Department, University of Jammu, Jammu, India\\
$^{96}$ Physics Program and International Institute for Sustainability with Knotted Chiral Meta Matter (SKCM2), Hiroshima University, Hiroshima, Japan\\
$^{97}$ Physikalisches Institut, Eberhard-Karls-Universit\"{a}t T\"{u}bingen, T\"{u}bingen, Germany\\
$^{98}$ Physikalisches Institut, Ruprecht-Karls-Universit\"{a}t Heidelberg, Heidelberg, Germany\\
$^{99}$ Physik Department, Technische Universit\"{a}t M\"{u}nchen, Munich, Germany\\
$^{100}$ Politecnico di Bari and Sezione INFN, Bari, Italy\\
$^{101}$ Research Division and ExtreMe Matter Institute EMMI, GSI Helmholtzzentrum f\"ur Schwerionenforschung GmbH, Darmstadt, Germany\\
$^{102}$ Saga University, Saga, Japan\\
$^{103}$ Saha Institute of Nuclear Physics, Homi Bhabha National Institute, Kolkata, India\\
$^{104}$ School of Physics and Astronomy, University of Birmingham, Birmingham, United Kingdom\\
$^{105}$ Secci\'{o}n F\'{\i}sica, Departamento de Ciencias, Pontificia Universidad Cat\'{o}lica del Per\'{u}, Lima, Peru\\
$^{106}$ Stefan Meyer Institut f\"{u}r Subatomare Physik (SMI), Vienna, Austria\\
$^{107}$ SUBATECH, IMT Atlantique, Nantes Universit\'{e}, CNRS-IN2P3, Nantes, France\\
$^{108}$ Sungkyunkwan University, Suwon City, Republic of Korea\\
$^{109}$ Suranaree University of Technology, Nakhon Ratchasima, Thailand\\
$^{110}$ Technical University of Ko\v{s}ice, Ko\v{s}ice, Slovak Republic\\
$^{111}$ The Henryk Niewodniczanski Institute of Nuclear Physics, Polish Academy of Sciences, Cracow, Poland\\
$^{112}$ The University of Texas at Austin, Austin, Texas, United States\\
$^{113}$ Universidad Aut\'{o}noma de Sinaloa, Culiac\'{a}n, Mexico\\
$^{114}$ Universidade de S\~{a}o Paulo (USP), S\~{a}o Paulo, Brazil\\
$^{115}$ Universidade Estadual de Campinas (UNICAMP), Campinas, Brazil\\
$^{116}$ Universidade Federal do ABC, Santo Andre, Brazil\\
$^{117}$ Universitatea Nationala de Stiinta si Tehnologie Politehnica Bucuresti, Bucharest, Romania\\
$^{118}$ University of Cape Town, Cape Town, South Africa\\
$^{119}$ University of Derby, Derby, United Kingdom\\
$^{120}$ University of Houston, Houston, Texas, United States\\
$^{121}$ University of Jyv\"{a}skyl\"{a}, Jyv\"{a}skyl\"{a}, Finland\\
$^{122}$ University of Kansas, Lawrence, Kansas, United States\\
$^{123}$ University of Liverpool, Liverpool, United Kingdom\\
$^{124}$ University of Science and Technology of China, Hefei, China\\
$^{125}$ University of South-Eastern Norway, Kongsberg, Norway\\
$^{126}$ University of Tennessee, Knoxville, Tennessee, United States\\
$^{127}$ University of the Witwatersrand, Johannesburg, South Africa\\
$^{128}$ University of Tokyo, Tokyo, Japan\\
$^{129}$ University of Tsukuba, Tsukuba, Japan\\
$^{130}$ Universit\"{a}t M\"{u}nster, Institut f\"{u}r Kernphysik, M\"{u}nster, Germany\\
$^{131}$ Universit\'{e} Clermont Auvergne, CNRS/IN2P3, LPC, Clermont-Ferrand, France\\
$^{132}$ Universit\'{e} de Lyon, CNRS/IN2P3, Institut de Physique des 2 Infinis de Lyon, Lyon, France\\
$^{133}$ Universit\'{e} de Strasbourg, CNRS, IPHC UMR 7178, F-67000 Strasbourg, France, Strasbourg, France\\
$^{134}$ Universit\'{e} Paris-Saclay, Centre d'Etudes de Saclay (CEA), IRFU, D\'{e}partment de Physique Nucl\'{e}aire (DPhN), Saclay, France\\
$^{135}$ Universit\'{e}  Paris-Saclay, CNRS/IN2P3, IJCLab, Orsay, France\\
$^{136}$ Universit\`{a} degli Studi di Foggia, Foggia, Italy\\
$^{137}$ Universit\`{a} del Piemonte Orientale, Vercelli, Italy\\
$^{138}$ Universit\`{a} di Brescia, Brescia, Italy\\
$^{139}$ Variable Energy Cyclotron Centre, Homi Bhabha National Institute, Kolkata, India\\
$^{140}$ Warsaw University of Technology, Warsaw, Poland\\
$^{141}$ Wayne State University, Detroit, Michigan, United States\\
$^{142}$ Yale University, New Haven, Connecticut, United States\\
$^{143}$ Yonsei University, Seoul, Republic of Korea\\
$^{144}$ Yukawa Institute for Theoretical Physics, Kyoto University, Kyoto, Japan\\
$^{145}$  Zentrum  f\"{u}r Technologie und Transfer (ZTT), Worms, Germany\\
$^{146}$ Affiliated with an institute covered by a cooperation agreement with CERN\\
$^{147}$ Affiliated with an international laboratory covered by a cooperation agreement with CERN.\\

\end{flushleft} 